\documentclass[12pt]{article}
\usepackage{amsfonts}
\usepackage{amssymb}
\usepackage{graphics,psboxit,amsmath}
\usepackage{subfigure}
\usepackage{graphicx}
\usepackage{verbatim}
\usepackage{epic}

\def\hybrid{\topmargin 0pt \oddsidemargin 0pt 
        \headheight 0pt \headsep 0pt
        \textwidth 16,5cm 
        \textheight 23cm 
        \marginparwidth .875in
        \parskip 5pt plus 1pt \jot = 1.5ex}

\hybrid

\catcode`\@=11

\def\marginnote#1{}
\newcount\hour
\newcount\minute
\newtoks\amorpm
\hour=\time\divide\hour by60
\minute=\time{\multiply\hour by60 \global\advance\minute by-\hour}
\edef\standardtime{{\ifnum\hour<12 \global\amorpm={am}%
        \else\global\amorpm={pm}\advance\hour by-12 \fi
        \ifnum\hour=0 \hour=12 \fi
        \number\hour:\ifnum\minute<10 0\fi\number\minute\the\amorpm}}
\edef\militarytime{\number\hour:\ifnum\minute<10 0\fi\number\minute}

\def\draftlabel#1{{\@bsphack\if@filesw {\let\thepage\relax
   \xdef\@gtempa{\write\@auxout{\string
      \newlabel{#1}{{\@currentlabel}{\thepage}}}}}\@gtempa
   \if@nobreak \ifvmode\nobreak\fi\fi\fi\@esphack}
        \gdef\@eqnlabel{#1}}
\def\@eqnlabel{}
\def\@vacuum{}
\def\draftmarginnote#1{\marginpar{\raggedright\scriptsize\tt#1}}

\def\draft{\oddsidemargin -.5truein
        \def\@oddfoot{\sl preliminary draft \hfil
        \rm\thepage\hfil\sl\today\quad\militarytime}
        \let\@evenfoot\@oddfoot \overfullrule 3pt
        \let\label=\draftlabel
        \let\marginnote=\draftmarginnote
   \def\@eqnnum{(\theequation)\rlap{\kern\marginparsep\tt\@eqnlabel}%
\global\let\@eqnlabel\@vacuum} }

\def\draft2{
        \def\@oddfoot{\sl preliminary draft \hfil
        \rm\thepage\hfil\sl\today\quad\militarytime}
        \let\@evenfoot\@oddfoot \overfullrule 3pt
        \let\label=\draftlabel
        \let\marginnote=\draftmarginnote
   \def\@eqnnum{(\theequation)\rlap{\kern\marginparsep\tt\@eqnlabel}%
\global\let\@eqnlabel\@vacuum} }

\def\preprint{\twocolumn\sloppy\flushbottom\parindent 2em
        \leftmargini 2em\leftmarginv .5em\leftmarginvi .5em
        \oddsidemargin -.5in \evensidemargin -.5in
        \columnsep .4in \footheight 0pt
        \textwidth 10.in \topmargin -.4in
        \headheight 12pt \topskip .4in
        \textheight 6.9in \footskip 0pt
        \def\@oddhead{\thepage\hfil\addtocounter{page}{1}\thepage}
        \let\@evenhead\@oddhead \def\@oddfoot{} \def\@evenfoot{} }

\def\numberbysection{\@addtoreset{equation}{section}
        \def\theequation{\thesection.\arabic{equation}}}

\def\underline#1{\relax\ifmmode\@@underline#1\else
        $\@@underline{\hbox{#1}}$\relax\fi}

\def\titlepage{\@restonecolfalse\if@twocolumn\@restonecoltrue\onecolumn
     \else \newpage \fi \thispagestyle{empty}\c@page\z@
        \def\thefootnote{\fnsymbol{footnote}} }

\def\endtitlepage{\if@restonecol\twocolumn \else \newpage \fi
        \def\thefootnote{\arabic{footnote}}
        \setcounter{footnote}{0}} 

\catcode`@=12
\relax

\def\figcap{\section*{Figure Captions\markboth
        {FIGURECAPTIONS}{FIGURECAPTIONS}}\list
        {Figure \arabic{enumi}:\hfill}{\settowidth\labelwidth{Figure
999:}
        \leftmargin\labelwidth
        \advance\leftmargin\labelsep\usecounter{enumi}}}
 \relax
\def\tablecap{\section*{Table Captions\markboth
        {TABLECAPTIONS}{TABLECAPTIONS}}\list
        {Table \arabic{enumi}:\hfill}{\settowidth\labelwidth{Table
999:}
        \leftmargin\labelwidth
        \advance\leftmargin\labelsep\usecounter{enumi}}}
 \relax
\def\reflist{\section*{References\markboth
        {REFLIST}{REFLIST}}\list
        {[\arabic{enumi}]\hfill}{\settowidth\labelwidth{[999]}
        \leftmargin\labelwidth
        \advance\leftmargin\labelsep\usecounter{enumi}}}
 \relax
\makeatletter
\newcounter{pubctr}
\def\publist{\@ifnextchar[{\@publist}{\@@publist}}
\def\@publist[#1]{\list
        {[\arabic{pubctr}]\hfill}{\settowidth\labelwidth{[999]}
        \leftmargin\labelwidth
        \advance\leftmargin\labelsep
        \@nmbrlisttrue\def\@listctr{pubctr}
        \setcounter{pubctr}{#1}\addtocounter{pubctr}{-1}}}
\def\@@publist{\list
        {[\arabic{pubctr}]\hfill}{\settowidth\labelwidth{[999]}
        \leftmargin\labelwidth
        \advance\leftmargin\labelsep
        \@nmbrlisttrue\def\@listctr{pubctr}}}
 \relax
\makeatother

\def\ba{\begin{equation}}
\def\ea{\end{equation}}

\def\no{\noindent}

\def\IR{\relax{\rm I\kern-.18em R}}

\begin{document}

\renewcommand{\theequation}{\thesection.\arabic{equation}}
\csname @addtoreset\endcsname{equation}{section}

\newcommand{\eqn}[1]{(\ref{#1})}
\newcommand{\be}{\begin{eqnarray}}
\newcommand{\ee}{\end{eqnarray}}
\newcommand{\non}{\nonumber}
\begin{titlepage}
\strut\hfill
\begin{center}

\vskip -1 cm


\vskip 2 cm

{\Large \bf  The 4-dimensional Taub string}

{\bf Nikolaos A. Batakis}

\vskip 0.2in

Department of Physics, University of Ioannina, \\
45110 Ioannina,  Greece\\
{\footnotesize{\tt nbatakis@uoi.gr}}\\

\end{center}

\vskip .4in

\centerline{\bf Abstract}

\no 
The prototype of a Taub  string  is formed by successive
junctions of copies of  Taub's space $ {\cal T}$, joined  
at their null boundaries  $\Sigma$  to create the 
axially-symmetric  Bianchi-type-XI (with compact SL sections 
of  scale ${\rm L_o}$) vacuum \mbox{${\cal B}^4_{\rm T}=
\dots\vee {\cal T}\vee {\cal T}\vee {\cal T}\vee\dots\;$},
which is a {\em proper} one, namely a stable, non-singular, 
geodesically and globally fit solution of  Einstein's  
vacuum equations without torsion and without a cosmological 
constant. Each  ${\cal T}$ contributes to ${\cal B}^4_{\rm T}$ 
with its entire life-span  as a quantum  of  time 
$\delta t\sim{\rm L_ o}$  between two consecutive $\Sigma$. 
The latter propagate as shock-wave fronts under  string 
tension of  Planck-scale strength $\kappa_{\rm o}$. The incurring  
dynamics entails  stability and the foundation of hierarchy in 
${\cal B}^4_{\rm T}$. Appropriate averaging of this dynamics
generates  {\em effective} stress-energy content and torsion 
in a static $\bar{\cal B}^4_{\rm T}$ vacuum. With the latter 
as ground state, excitations thereof must involve two new  
independent scales, $\kappa$ and ${\rm L_1}\gg{\rm L_o}$,
in addition to $\kappa_{\rm o},\;{\rm L_o}$. Elemental finite 
Taub  strings and variant vacua, including the 
${\rm L_o}\gg\kappa_{\rm o}$  cosmological case,  
are also discussed.

\vfill
\no


\end{titlepage}
\vfill
\eject

\no

\section{Introduction}
The elegance in current geometrical settings, mitigated as it is
by fundamental problems  on stability and hierarchy,
offers strong motivation 
for the search of new perspectives in context. 
At the accordingly  fundamental level
of a spacetime manifold, we will resort to the notions 
of a  {\em proper vacuum}\footnote
{Namely a  solution of  Einstein's  vacuum equations
without torsion and without a cosmological constant which is
mathematically and physically acceptable, namely
non-singular,  stable, geodesically  and
globally fit, especially with no closed TL curves
due to vorticity, chaoticity or topology beyond Planck sale.}
${\cal V}$ and of an {\em effective proper vacuum}\footnote
{A non-static ${\cal V}$ with compact SL sections
could, under time averaging, descend  to a static $\bar{\cal V}$
with acquired features, such as effective energy-momentum and 
torsion content, as  relics of the original dynamics.} $\bar{\cal V}$, aiming to
search beyond   Minkowski's ${\cal M}^4_{\rm o}$ (which,  modulo  pp-waves, is 
the only 4D proper  vacuum  with open SL sections)
to find all possible 4D proper vacua  with {\em compact} SL
sections. 

To do that, one must exhaust the classification of 
homogeneous spacetimes \cite{helga}, \cite{rs}, \cite{duff}, to
come-up with a  generic Bianchi-type IX, the ${\cal B}^4_{\rm IX}$, 
as the only such candidate. 
Its 3D orthochronous  sections, of any scale ${\rm L_o}$,
exhibit the same dynamics of $SU(2)$ acting transitively
on its own group manifold, the homogeneous $S^3$. The minimally 
symmetric case,  categorized as  ($3\!+\!0$) for 
its three transitive Killing vectors, is, of course,
Misner's  mixmaster\footnote 
{The term relates to the non-linear 
mixmaster dynamics, originally proposed
as a homogenization and isotropization mechanism for
the study of horizons and cosmic mixing
in relativistic cosmology.}
${\cal B}^4_{\rm M}$  \cite{misner}. 
Probably non-singular, ${\cal B}^4_{\rm M}$ is certainly  beyond analytic treatment  except for  exact
or effective symmetries, such as those due to chaotic mixing  
over scales sufficiently larger 
than ${\rm L_o}$. The  maximally-symmetric $(3\!+\!3)$ case 
is incompatible with the vacuum equations
and the $(3\!+\!2)$ case cannot exist either (the two generators  
cannot form a subgroup of isometries). 
The remaining $(3\!+\!1)$ case is
Taub's  axially-symmetric solution  ${\cal T}$ \cite{taub}, often treated as if it were 
singular, which it is not; ${\cal T}$
does possess an initial and a final {\em physical} singularity,
namely its pair of  null  boundaries $\Sigma$, but
at such singularities (in contrast to mathematical ones)  
the volume element and the entire Riemann tensor must
remain finite.  Regardless of coordinate failure on the $\Sigma$,  
the problem  is with  extending the 
geodesics beyond those 
boundaries: it is a pathogeny to be
cured, otherwise ${\cal T}$ will remain only part of a missing whole.
The Taub-NUT construction 
does cure it \cite{mistaub} but  still fades as  
a  proper vacuum on certain global aspects, and so does
any finite number of ${\cal T}$ joined  in a closed or open string
formation\footnote
{The  Taub-NUT space  is  formed by a ${\cal T}$ sandwiched 
via two $\Sigma$ junctions (a pair of `Misner bridges') between two NUT 
spaces as NUT$\vee {\cal T}\vee$NUT. 
In the present construction we  use only Taub copies in
consecutive junctions, indefinitely, 
as $\cdots\vee {\cal T} \vee {\cal T}\vee {\cal T}\vee\cdots$,
for a likewise non-singular result.}.
We have thus been led to the axially symmetric
proper 4D vacuum
\be
{\cal B}^4_{\rm T} = \bigvee^{n=+\infty}_{n=-\infty}{\cal T}_n
=\cdots\vee{\cal T}_{n-1}
\vee{\cal T}_{n}
\vee{\cal T}_{n+1}\cdots,\;\;\;\; n\in Z 
\ ,
\label{taub}
\ee
as prototype of a Taub string, essentially unique, given the elusive nature of
Misner's ${\cal B}^4_{\rm M}$.
The  integer  $n$ enumerates
the ${\cal T}_n$
and their boundaries $\Sigma_n$  (null squashed $S^3$)  at the junctions.
The periodic  metric of ${\cal B}^4_{\rm T}$
is expressible in terms of the $b=b(u)$, $c=c(u)$ radii (cf. also Fig.\ref{fig}
in section 2)
as functions of $u$, which is a null
coordinate,  namely with $(\partial_u)^2=0$,
and of the  $\theta,\phi,\psi$ coordinates on the homogeneous $du=0$
squashed-$S^3$ hypersurfaces as
\be
ds^2=-2{\rm L_o}du(d\psi\!+\!\cos\theta d\phi)+
{\rm L_o}^2\left(b^2[(d\theta)^2+\sin^2\theta (d\phi)^2]
+c^2(d\psi+\cos\theta d\phi)^2\right)
\  .
\label{tmc}
\ee
The   $\theta,\phi,\psi$  also define  
the  $SU(2)$ left-invariant  $\ell^\mu$ ($\ell^0\!\!=\!\!du$, 
$\ell^1/{\rm L_o}\!\!=\!\!\cos\psi d\theta+\sin\theta\sin\psi d\phi$,
$\ell^2/{\rm L_o}\!\!=\!-\sin\psi d\theta+\sin\theta\cos\psi d\phi$,
$\ell^3/{\rm L_o}\!\!=\!\cos\theta d\phi+d\psi$ for
$d\ell^i\!=\!-\frac{1}{2{\rm L_o}}\epsilon^i_{jk}\ell^j\wedge\ell^k$)
with dual $L_\mu$ ($L_0\!\!=\!\!\partial_u$, $L_i$),  and
$g_{03}\!=\! -1,\; g_{11}\!=\! g_{22}\!=\! b^2,\; g_{33}\!=\! c^2$ 
in a non-holonomic equivalent of (\ref{tmc}) as
\be 
ds^2=-2du(\ell^3)+ 
b^2\left[(\ell^1)^2+(\ell^2)^2\right]+c^2(\ell^3)^2.
\ 
\label{tm}
\ee
As we will see, in-between the  boundaries 
$\Sigma_n$, $\Sigma_{n+1}$ of ${\cal T}_n$,
the  $du=0$  sections in (\ref{tm}) remain  SL
while being transported by $\partial_u$ but 
they become momentarily null when they arrive at (and identify with) 
$\Sigma_{n+1}$. The latter is the final  boundary of ${\cal T}_n$ and
simultaneously the initial  of ${\cal T}_{n+1}$.
Each  ${\cal T}$ contributes to ${\cal B}^4_{\rm T}$ 
with its entire life-span  as a quantum  of  time 
$\delta t\sim{\rm L_ o}$  between two consecutive  
$\Sigma$, which propagate as   
shock-wave fronts under  string tension
of (roughly) Planck-scale $\kappa_{\rm o}$ strength. The incurring  dynamics
entails  stability and the foundation  
of hierarchy in ${\cal B}^4_{\rm T}$.
Appropriate averaging of this dynamics
generates  {\em effective} stress-energy content, possibly from effective torsion 
in a static $\bar{\cal B}^4_{\rm T}$ vacuum.
We'll also discuss variant vacua including finite  
Taub  strings and the cosmological ${\rm L_o}\gg\kappa_{\rm o}$ case.

\no
\underline {Notation \& conventions:} 
The metric in $ds^2\!=\!g_{\mu\nu}\ell^\mu\ell^\nu$, with
$dg_{\mu\nu}= \Gamma_{\mu\nu}+\Gamma_{\nu\mu}$, simplifies to
$\eta_{\mu\nu}\!=\!(-1,+1,+1,+1)$ in 
orthonormal Cartan coframes $\theta^\mu=\theta^\mu_\nu\ell^\nu$ with
dual $\Theta_\mu$; the
general connection $\gamma^\mu_{\;\;\nu}$ and the Christoffel 
$\Gamma^\mu_{\;\;\nu}=\Gamma^\mu_{\;\;\nu\rho}\theta^\rho$
(with covariant derivatives ${\cal D}$, $D$,  respectively)  
are antisymmetric in $\mu,\nu$ just like the
contorsion tensor-valued 1-form $K^\mu_{\;\;\nu}$ in
\be
\gamma^\mu_{\;\;\nu}=\Gamma^\mu_{\;\;\nu} +K^\mu_{\;\;\nu} ,\;\;\;\;\;\;
D\theta^\mu:=d\theta^\mu+\Gamma^\mu_{\;\;\nu}\wedge\theta^\nu\equiv 0,\;\;\;\;\;
{\cal D}\Theta_\mu=d\Theta_\mu-\Gamma^\nu_{\;\;\mu}{\cal L}_\nu\equiv 0
\ .
\label{conect}
\ee
The general curvature ${\cal R}^\mu_{\;\;\nu}$
includes its Riemannian  part $R^\mu_{\;\;\nu}:=
d\Gamma^\mu_{\;\;\nu}+\Gamma^\mu_{\;\;\rho}\wedge
\Gamma^\rho_{\;\;\nu}$, with 
${W}^\mu_{\;\;\nu\rho\sigma}$  and
$R_{\mu\nu}\!=\!R^\rho_{\;\mu\rho\nu}$ 
the Weyl and Ricci tensors. 
Cartan's first and second structure equations 
involve the general curvature ${\cal R}^\mu_{\;\;\nu}$
and  the torsion tensor-valued 2-form $T^\mu$  as  \cite{traut}
\be
{\cal R}^\mu_{\;\;\nu}:&=&d\gamma^\mu_{\;\;\nu}+\gamma^
\mu_{\;\;\rho}\wedge\gamma^\rho_{\;\;\nu}=R^\mu_{\;\;\nu}+
DK^\mu_{\;\;\nu}+K^\mu_{\;\;\rho}\wedge K^\rho_{\;\;\nu}=
\frac{1}{2}{\cal R}^\mu_{\;\;\nu\rho\sigma}\theta^\rho\wedge \theta^\sigma ,
\ 
\label{riemann}\\
T^\mu:&=&{\cal D} \theta^\mu=d\theta^\mu+\gamma^\mu{\!}_\nu\wedge \theta^\nu =
K^\mu{\!}_\nu\theta^\nu=
\frac{1}{2}T^\mu_{\;\;\rho\sigma}\theta^\rho\wedge \theta^\sigma 
\label{torsion}
\ . 
\ee
The  geodesic and Killing (by the ${\cal L}_\Xi$ Lie derivative) equations
are employed as  
\be
U=U^\mu \Theta_\mu &:&  \;\;\;\;
\nabla_U {U}=0\;\longrightarrow\;\;\;DU^\mu (U)=0\;\; {\rm [extending\;to]}\;\;
{\cal D}U^\mu (U)=0, 
\label{geo}
\\
\Xi =\Xi^\mu \Theta_\mu &:& \;\;\;\;
{\cal L}_\Xi {\theta^\mu}=d\Xi^\mu+(\Gamma^\mu_{\;\rho\nu}
-\Gamma^\mu_{\;\nu\rho})\Xi ^\rho\theta^\nu
\; \longrightarrow\;\;\;D\Xi^\mu=\Gamma^\mu_{\;\nu\rho}\Xi ^\rho\theta^\nu,
\
\label{kill}
\ee
with $U\cdot U\!=\!U_\mu U^\mu\!=\!\varsigma=0,\mp 1 $ for the null, TL or SL cases. 

\no
\section {The ${\cal B}^4_{\rm T}$ proper vacuum as a Taub string}
 
\begin{figure}
\begin{center}
\includegraphics[width=10cm]{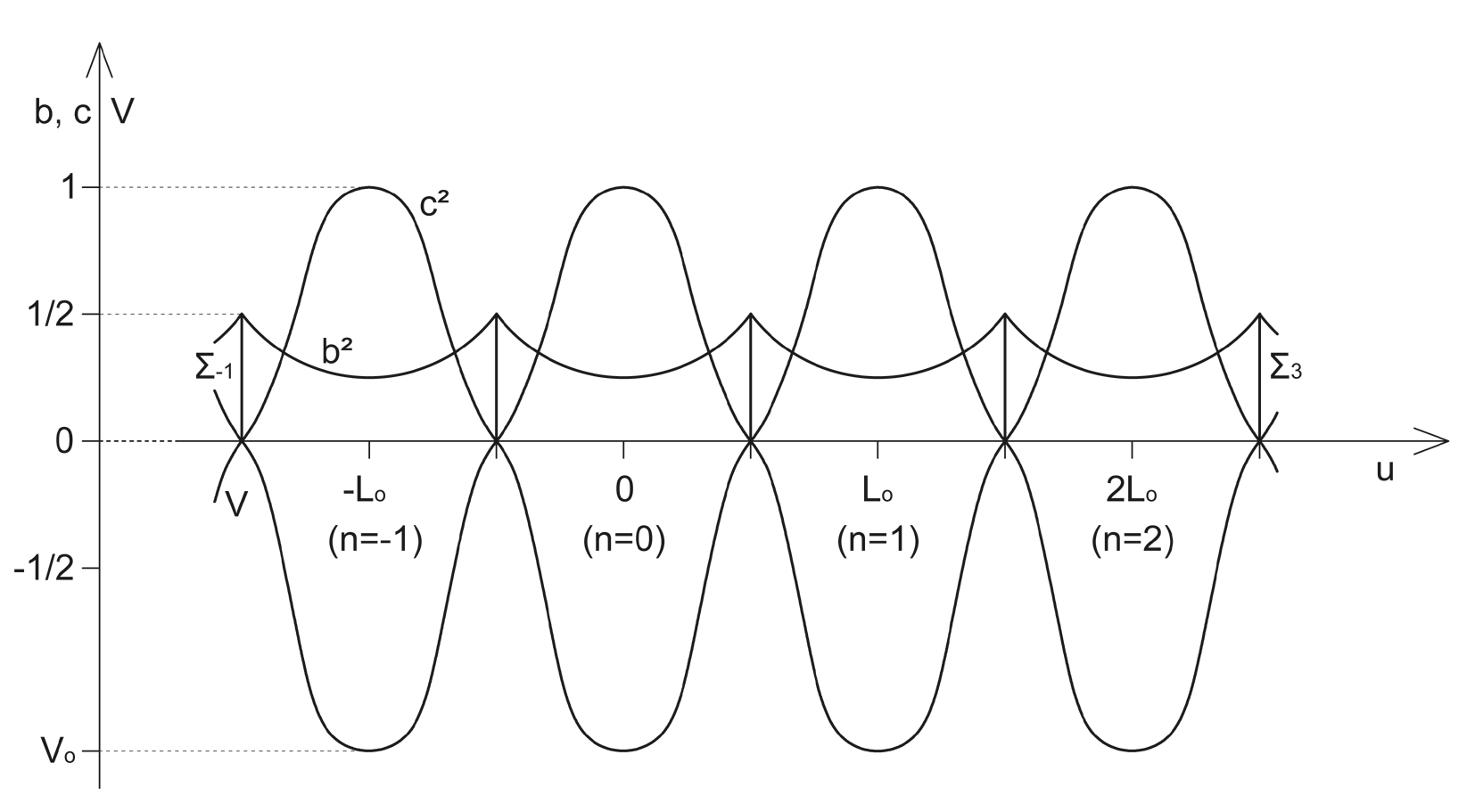}
\caption{The Taub string  ${\cal B}^4_{\rm T}$,
depicted in terms of its radii in the  metric 
(\ref{tmc}), with the $b^2$, $c^2$  of (\ref{solution})
in units of ${\rm  L_o}$ (with $B=2$),
along  the potential wells  $V=V(u)$.
The short vertical lines
represent  the propagating null squashed-$S^3$ junction hypersurfaces 
$\Sigma_{-1}$ to $\Sigma_3$.}
\label{fig}\;
\end{center} 
\end{figure}

Taub's general  solution for ${\cal T}$, to be expressed  in terms of the 
$b=b(u)$, $c=c(u)$ functions in (\ref{tm}), needs a subtle 
refinement for compatibility at every junction
across the $\Sigma_n$ boundaries in (\ref{taub})
and thence for the entire  ${\cal B}^4_{\rm T}$.
In addition to ($\ell^\mu$, $L_\nu$), we will also employ two different
sets of {\em orthonormal} Cartan  frames in  ${\cal B}^4_{\rm T}$,
the ($\theta^\mu$, $\Theta_\nu$) and the ($e^\mu$, $E_\nu$), with
 \be
\theta^0&=&c^{-1}du,\;\;\theta^1=b\ell^1,
\;\;\theta^2 = b\ell^2, \;\; \theta^3 = c\ell^3-c^{-1}du\;;\;
\label{frames}
\\
\Theta_0 &=&c\partial_u+c^{-1}L_3,\;\; 
\Theta_1 =b^{-1}L_1,\;\; \Theta_2 =b^{-1}L_2, \;\;              
\Theta_3=c^{-1}L_3,                           
\nonumber
\ee
\be
e^0\!&=&\!\frac{1}{\sqrt{2}}\left[du+(1-\frac{c^2}{2})\ell^3\right]\!\!,
\;e^1 = b\ell^1, \;e^2 = b\ell^2, \; 
e^3=\frac{1}{\sqrt{2}}\left[du-(1+\frac{c^2}{2})\ell^3\right],
\label{Frames}
\\
E_0\!&=&\!\frac{1}{\sqrt{2}}\left[(1+\frac{c^2}{2})\partial_u+L_3\right]\!\!,\;
E_1= b^{-1}L_1,\;E_2\!=\! b^{-1}L_2,\;
E_3=\frac{1}{\sqrt{2}}\left[(1-\frac{c^2}{2})\partial_u-L_3\right]\!\!,
\nonumber
\ee
which are quite useful in spite of the failure of 
($\theta^\mu$, $\Theta_\nu$) on  $\Sigma_n$
and the non-holonomic time in  ($e^\mu$, $E_\nu$).
The Christoffel 1-forms $\Gamma^0_{\;\;1}=c(\ln b)\dot{}\,\theta^1$,
$\Gamma^0_{\;\;3}=\dot{c}\,\theta^3$,  etc.,
with a dot for $d/du$,  provide the  
Riemann and Weyl tensors (finite everywhere) 
in the ($\theta^\mu$, $\Theta_\nu$) frames  as
\be
R^0_{\;\;101}=R^0_{\;\;202}&=&c^2(\ln b){\,}\ddot{}
+c^2(\ln b)\dot{}^{\;2}+c\dot{c}(\ln b)\dot{}\; ,\;\;\;\;\;\;\;\;\;
R^0_{\;\;303} = c\ddot{c}+\dot{c}^2\ \label{riemann1}
\\
R^2_{\;\;323}=R^3_{\;131}&=& 
 c\dot{c}(\ln b)\dot{}+\frac{c^2}{4{\rm L_o}^2b^4}\;,\;\;\;\;\;\;\; 
R^1_{\;\;212} = c^2(\ln b)\dot{}^{\;2}+\frac{4b^2-3c^4}{4{\rm L_o}^2b^4}
\label{riemann2}\\
R^0_{\;\;123}=R^0_{\;\;231}&=&-\frac{1}{2}R^0_{\;\;312}=W^0_{\;\;123}=
W^0_{\;\;231}=-\frac{1}{2}W^0_{\;\;312}
=\frac{1}{4{\rm L_o}}\left(\frac{c^2}{b^2}\right)^{\!\!\!\cdot}
\label{riemann3}\;.
\
\ee
In addition to the basic scale ${\rm L_o}$ from the frames,
Taub's general solution of $R_{\mu\nu}=0$ has 
two independent constant parameters, $B>0$ and $N\in\IR$,
essentially from time scaling and translational invariance, respectively. 
However,  that solution may also be expressed as
\be
b^2 = \frac{1}{2B}\left(1+4\left(\frac{u}{{\rm L'_o}}\right)^2\right),\;\;
c^2 =\frac{2}{B}\cdot\frac{1-4\left(u/{\rm  L'_o}\right)^2}
{1+4\left(u/{\rm L'_o}\right)^2}\;,\;\;\;\;\;\;\;\;\;\;\;
\left[ {\rm L'_o}=\frac{2{\rm L_o}}{B}\right],
\label{solution}
\ 
\ee
and still remain the general  one, in spite of the absence of $N$.  
The latter is  hidden in the allowed
$u\rightarrow u+N{\rm L'_o}$  translations 
(with $N\in\IR$, ${\rm  L'_o}=2{\rm  L_o}/B$), under which (\ref{solution}) is
{\em not} form invariant but  remains a solution of $R_{\mu\nu}=0$.
This will be utilized as a tool to  reform the 
general Taub solution  ({\em with} the $N\in\IR$) in (\ref{solution}), 
to also being, simultaneously, the
${\rm  L'_o}$-periodic 
solution for ${\cal B}^4_{\rm T}$ in (\ref{taub}). 
However, in the second interpretation,  
$N\in\IR$ no-longer exists as an independent parameter because,
restricted to its integer values as $N\in Z$, 
it has been consumed as a `summation' index $n$ in (\ref{taub}),
as  depicted in  Fig.\ref{fig}.
Thus, the original  translational invariance,
with (\ref{solution}) as the Taub general solution, 
has been traded for periodicity 
in (\ref{taub}),  now viewed as the ${\cal B}^4_{\rm T}$ solution. The
ensuing quantization of the time coordinate is crucially involved in 
stability and hierarchy aspects, interrelated
via string tension of a
$\kappa^2_{\rm o}$ strength, as we will see.
We still have, of course, 
$(\partial_u)^2=0$, $(\partial_u)\cdot(L_3)^2=-1$,
$(L_3)^2=c^2$ so, by the latter, we verify that  $L_3$ 
turns momentarily null and orthogonal to the $\Sigma_n$ at 
the $\pm {\rm  L'_o}/2$  roots of $c^2=0$,
as follow from  (\ref{solution}). 
The entire life-time span of ${\cal T}$, as it extends in-between
these roots, constitutes 
an elemental `quantum of time' contribution 
to ${\cal B}^4_{\rm T}$. Its proper-time duration $\delta t$
is actually calculable along
TL geodesics in terms of the $u$ coordinate in the 
\mbox{$-{\rm  L'_o}/2\leq u\leq{\rm  L'_o}/2$} interval.
The latter, now realized as the $n=0$ one in (\ref{taub}),
extends periodically 
as  \mbox{$[n-{\rm  L'_o}/2,\; n+{\rm  L'_o}/2]$} for any $n\in Z$,
as also  depicted in Fig. \ref{fig}.
At the extremes of these intervals, where the 
physical singularities  $\Sigma_n$  of the ${\cal T}_n$ are located,
the volume element never vanishes and
the components  of $R^\mu_{\;\;\nu}$ never diverge.

Our less-than-$C^2$  junctions at the $\Sigma_{n}$   
boundaries in ${\cal B}^4_{\rm T}$ of (\ref{taub})  
may  (and  they do) carry surface tensions, calculable from 
$C^0$-junction compatibility conditions. 
The normal to the propagating $du=0$  hypersurfaces
is the TL unit vector  ${\cal N}=\Theta_0$, but
these hypersurfaces become null  at $c=0$, 
where they reach (and identify with) the $\Sigma$.
The normal to the latter is the (momentarily null there) $L_3$ vector,
so the extrinsic curvature vector-valued 1-form is ${\cal K}=-dL_3$.
The discontinuity (jump) $\delta {\cal K}$
across every  $\Sigma$ reveals a stress-energy layer thereon,
which creates  string tension across every  $\Sigma$ 
along  ${\cal B}^4_{\rm T}$. To see that, we can start
the computation of the stress-energy tensor $S_{\mu\nu}$ layers
with  ${\cal N}=L_3=c\Theta_3$  in the ($\theta^\mu$, $\Theta_\nu$)
frame  of (\ref{frames}) and, 
after cancellation of the anomalies 
at $c=0$, switch back to the ($\ell^\mu$, $L_\nu$).
We thus find  that the discontinuities $\delta {\cal K}=-\delta d(c\Theta_3)$ 
occur only on the (0,3) plane  across  $\Sigma$  and only
as $\delta{\cal K}^0_{\;\;3} =\delta {\cal K}^3_{\;\;0}=-\frac{4}{\rm L_o}(c^2)\dot{}$.
The full result for the $S_{\mu\nu}$ layers, expressed in either of the ($\ell^\mu$, $L_\nu$),
($e^\mu$, $E_\nu$) frames (the $\theta^\mu$, $\Theta_\nu$ fail on 
$\Sigma_n$),  involves the only non-vanishing 
\be
S_{00}=S_{33}=
+\frac{2B}{{\rm L_o}^2\kappa^2_{\rm o}}\;
\sum_{n=-\infty}^{n=\infty}\delta\left(u-(n+\frac{1}{2}){\rm L'_o}\right),
\;\;\;\;n\in Z
\ ,
\label{stress}
\ee
components of a $S_{\mu\nu}\sim{\rm diag}(1,0,0,1)$ 
layer of  Planck-scale
$\kappa^2_{\rm o}$ strength, traceless and with correct content and sign for string tension
orthogonal to the  $L_1,\;L_2$ directions.
This self-consistent (with no external sources) geometric disturbance (which cannot
be  a pp-wave) occurs  across the  compact null interface  $\Sigma$
and involves the latter as
a gravitational  shock-wave front \cite{c-b}. Thus, the  stability of the  dynamics
in-between the $\Sigma_n$ is  being extended across them and
over the entire  ${\cal B}^4_{\rm T}$ manifold\footnote
{The synergy of  breathing-mode dynamics  with the $V(u)$
potential and the propagating  shock-wave fronts 
for ${\cal B}^4_{\rm T}$, as shown  in Fig.\ref{fig},
may  offer a distant
resemblance to Planck-scale mixmaster dynamics in ${\cal B}^4_{\rm M}$.}
(cf. also discussion in the last section).

For the  $U=U^\mu\Theta_\mu$ tangent to geodesics  in ${\cal B}^4_{\rm T}$,
with $u=u(t)$, $\chi=\chi(t)$ in terms of the proper-time parameter $t$
and $du/dt=1/\dot t$
($P_\perp$, $E$ are constants of motion),  we find
\be
U^0=\frac{1}{c\dot t} ,\;\; U^1=P_\perp \frac{\cos \chi}{b} ,
\;\; U^2=-P_\perp \frac{\sin \chi}{b} ,
\;\;U^3= \frac{\sqrt{2E}}{c},
\ 
\label{u}
\ee
plus two first integrals  expressible in terms of the given $v=v(u)$ function as 
\be
\dot \chi &=& \frac{c^2-b^2(1+\cosh v) }{{\rm L_o}b^2c^2\cosh v}\;, \;\;\;\;\;\;\;\;\;\;
\;\;\;\;\;\;\;
\;\;\;\;\;\;\;\;\;\sinh^2v: =\frac{1}{2E}
 \left(\frac{ P_\perp^2}{b^2}-\varsigma\right) c^2\;, 
\label{fi}
\\
\dot t &=&\frac{1}{\sqrt{2E}\cosh v}\;\;\;\longleftrightarrow\;\;\;
\frac{1}{2}\left(\frac{du}{dt}\right)^{\!2}\! +V(u) = E\;,\;\;\;\;\;\;\;\;\;\; V(u) = -E\sinh^2v\;.
\ 
\label{e}
\ee
The so-emerging $n=0$ potential-well function of  $V=V(u)$,
as shown in Fig.\ref{fig}, is a known  stability aspect
of the dynamics  within ${\cal T}$. It has a minimum 
of  $-(2P_\perp^2+1/B)$ at $u=0$,  with
s-c ({\em supremum}-case)
value of $V_o=-1/B$ if $P_\perp=0$, and  TL geodesics (with $E<0$)
trapped within the wells of all  $n$.
The latter aspect,  not quite the same as geodesic inextendibility
(especially if it is involved at Planck scale),  may survive our 
${\cal B}^4_{\rm T}$ completion
of ${\cal T}$, as  in the Taub-NUT case. 
In the general dynamics
from  (\ref{e}), all  $E>0$ geodesics  
in the singularity-free geometry of ${\cal B}^4_{\rm T}$
propagate  freely, 
except for an instantaneous  impact at the junctions, due to
stresses on the $\Sigma$ boundaries. Nevertheless,
we do have  peculiar behavior in the case of s-c geodesics.
They certainly  form a marginal  subclass (due to the
vigorous $P_\perp=0$ requirement), but the  dynamics can be easily
integrated in that case to the elegant result
\be
U^0=\sqrt{1+\frac{2E}{c^2}}=\gamma,\;\;\;U^3=\sqrt{\frac{2E}{c^2}}=
\beta\gamma \;\;\;\;\;\;\left[\gamma=
\frac{1}{\sqrt{1-\beta^2}}, \;\;\;\; \beta^2=\frac{2E}{c^2+2E}\right]
\ ,
\label{beta}
\ee
also expressed here in terms of the  
special-relativistic $\beta$ and $\gamma$ parameters  
on the $(\Theta_0 ,\Theta_3)$ plane. Expectedly,
all s-c geodesics  turn momentarily null as they cross
the $\Sigma$ boundaries, where $\beta\rightarrow 1$ 
as $c\rightarrow 0$, but  with  complete `loss of memory'  
of whatever $E>0$  value they 
had, due to its cancellation (an unexpected result we will return to).

\no
\section {The $\bar{\cal B}^4_{\rm T}$ Taub string with effective  torsion} 

{\em A priori}, ${\rm L_o}$  can be  equally-well identified with
Planck length or with Hubble radius,  but  in the former case the notion of 
a regime of scales `much larger' than ${\rm L_o}$  can be made 
physically and mathematically relevant.  
Our elemental `quantum-of-time' $\delta t\!\sim\!{\rm L_o}$,
equivalently taken as a $\delta u\!=\!{\rm L'_o}$, can then be formally treated 
as an  ordinary-calculus differential  $du$, allowing
the reduction of  ${\cal B}^4_{\rm T}$ to  its
{\em effective} $\bar{\cal B}^4_{\rm T}$ by 
a well-defined averaging  process: at this limit,
differences between
adjacent values of  $n\in Z$ are negligible, $n\in Z$ is treatable as a continuous $N\in \IR$
and summations over  $n$ as normalized integrations  over $u$, to give us
\footnote
{Clearly distinct from the classical  limit ${\rm L_o} \rightarrow 0$, the  
`de-quantization' limit  ${\rm L'_o}\rightarrow du$ 
equips the entire  process with calculus-approachable dynamics,
by which $\bar{\cal B}^4_{\rm T}$ can be fully treated as a differentiable manifold.}
\be
\overline{b^2}:=\frac{1}{\rm L'_o}\int_{-{\rm L'_o}/2}^{{\rm L'_o}/2}
b^2du=\frac{2}{3B}\;,\;\;\;\;
\overline{c^2}:=\frac{1}{\rm L'_o}\int_{-{\rm L'_o}/2}^{{\rm L'_o}/2}c^2du=\frac{\pi-2}{B}\;,
\ 
\label{avg}
\ee 
\be
\overline{(b^2)\dot{}}=0,\;\overline{(c^2)\dot{}}=0 ;\;\;\;\;
\overline{[(b^2)\dot{}\,]^2}=\frac{(\pi-2)B^2}{2{\rm L_o}^2} ,\;\dots\;;
\;\;\;\;\overline{(b^2)\,\ddot{}}=\frac{B}{{\rm L_o}^2} ,\;
\;\overline{(c^2)\,\ddot{}}=-\frac{2B}{{\rm L_o}^2}\;; \;\;\dots
\  
\label{avgd}
\ee
as results which follow from (\ref{solution}) etc. 
In spite of the  presence of ${\rm L_o}=B{\rm L'_o}/2$, these results  
hold only in the higher-scale differences of the  $\delta u>>{\rm L'_o}$ regime, 
and, in spite of the $\overline{(b^2)\dot{}}\!\!=\!\! 
\overline{(c^2)\dot{}}\!\!=\!\!0$ vanishings, remnants of
the original dynamics in ${\cal B}^4_{\rm T}$
effectively do survive
in $\bar{\cal B}^4_{\rm T}$.
To uncover them, we use  (\ref{avg}) in 
(\ref{tm}) to find the  metric of this {\em static}
$\bar{\cal B}^4_{\rm T}$ as \footnote
{The bar in $d\bar s^2$ and over quantities like $\bar b$, $\bar{\Gamma}^{\mu}_{\;\nu}$
or $\bar {\cal R}^\mu_{\;\nu}$ is only a marker associating them to
$\bar{\cal B}^4_{\rm T}$. For averaging we overline, as
with $\overline{b^2}$,
$\overline{[(b^2)\dot{}\,]^2}$,
so expressions  like  those involved in (\ref{bbb}) below are well defined.}
\be
d\bar s^2=-2(du)\ell^3+ 
\bar b^2[(\ell^1)^2+(\ell ^2)^2]+\bar c^2(\ell^3)^2,\;\;
\left[\bar g_{03}= -1,\;\bar g_{11}=\bar g_{22}=\frac{2}{3B},\;\bar g_{33}\!=\frac{\pi-2}{B}
\right]
\ 
\label{bartm}
\ee 
with
$\bar b^2=\overline{b^2}$ and $\bar c^2=\overline{c^2}$. 
All derivatives of $\bar b^2$, $\bar c^2$ vanish, of course, 
in sharp contrast  to the 
results in  (\ref{avgd}). We thus find 
(with $\bar a^2\!\!:=\bar c^2-\bar b^2$) 
\be
\bar{\Gamma}^{\mu}_{\;\;\nu}= \frac{1}{2{\rm L_o}\bar b^2}
\left[\begin{array}{cccc} 0&0&0&0\\
-\ell^2&0&-\ell^0\!\!-\!\!(\bar b^2-\bar a^2)\ell^3&\bar c^2\ell^2\\
\ell^1&\ell^0\!\!+\!\!(\bar b^2-\bar a^2)\ell^3&0&-\bar c^2\ell^1\\
0&-\bar b^2\ell^2&\bar b^2\ell^1&0
\end{array} \right],\;\;\;\;\;\bar a^2:=\bar c^2-\bar b^2,
\label{avriemann}
\ee
\be
\bar R^{\mu}_{\;\;\nu}= \frac{1}{(2{\rm L_o}\bar b^2)^2}
\left[\begin{array}{cccc} 0&0&0&0\\
\ell^1\!\!\wedge\!\!\left[\ell^0\!\!-\!\!\bar a^2\ell^3\right]
&0&-2\bar a^2\bar b^2\ell^1\ell^2&
-\bar c^2\ell^1\!\!\wedge\!\!\left[\ell^0\!\!-\!\!\bar a^2\ell^3\right]\\
\ell^2\!\!\wedge\!\!\left[\ell^0\!\!-\!\!\bar a^2\ell^3\right]
&-2\bar a^2\bar b^2\ell^2\ell^1&0&
-\bar c^2\ell^2\!\!\wedge\!\!\left[\ell^0\!\!-\!\!\bar a^2\ell^3\right]\\
0&-\bar b^2\left[\ell^0\!\!-\!\!\bar a^2\ell^3\right]\wedge\ell^1&
-\bar b^2\left[\ell^0\!\!-\!\!\bar a^2\ell^3\right]\wedge\ell^2&0
\end{array} \right],
\
\nonumber
\ee
By these results, which include an 
identically vanishing Weyl  tensor,
we conclude that, in comparison with the original ${\cal B}^4_{\rm T}$,
the shock-wave disturbance
has  disappeared, as expected, 
but the now conformally flat  $\bar{\cal B}^4_{\rm T}$
is no-longer Ricci flat. For a closer comparison  between the curvatures 
${R^\mu_{\;\;\nu}}$,  $\bar{R}^\mu_{\;\;\nu}$ of 
${\cal B}^4_{\rm T}$, $\bar{\cal B}^4_{\rm T}$
we must 
resort to the calculation of 
$\overline{R^\mu_{\;\;\nu}}$ directly from
\mbox{(\ref{riemann1}-\ref{riemann3})}, 
knowing beforehand that ${R^\mu_{\;\;\nu}}$
is already Ricci flat, hence it  will certainly have to remain so after the averaging.
Of course, due to the loss of Ricci flatness in $\bar{\cal B}^4_{\rm T}$,
we necessarily have
$\bar{R}^\mu_{\;\;\nu}\neq\overline{R^\mu_{\;\;\nu}}$, so the
emergence of effective stress-energy content
has already occurred in $\bar{\cal B}^4_{\rm T}$. 
If this content  is due to effective torsion $\bar{T}^\mu$,
that inequality can turn  into a precise relation 
between ${\cal B}^4_{\rm T}$ and $\bar{\cal B}^4_{\rm T}$.
Resorting to their basic observables, namely
their metrics  $g_{\mu\nu}$, $\bar g_{\mu\nu}$
and  their curvatures $ R^{\mu}_{\;\;\nu}$,  $\bar {\cal R}^{\mu}_{\;\;\nu}$ 
(now allowing for a general connection in $\bar{\cal B}^4_{\rm T}$),
all expressed in the same time-independent  left-invariant 
frame $\ell^\mu$, we can then demand that
 \be
\bar g_{\mu\nu}=\overline{g_{\mu\nu}}\;,\;\;\;\;\;\;
\bar {\cal R}^\mu_{\;\;\nu}:=
\bar {R}^\mu_{\;\;\nu}+\bar D\bar {K}^\mu_{\;\;\nu}+
\bar {K}^\mu_{\;\;\rho}\wedge\bar {K}^\rho_{\;\;\nu}=
 \overline{R^\mu_{\;\;\nu}}
\ , 
\label{bbb}
\ee
The first requirement is already satisfied in view of (\ref{bartm}). 
If $ \overline{R^\mu_{\;\;\nu}}$ is  known or calculable, we can
integrate the second  to find 
$\bar T^{\mu}=\bar{K}^{\mu}_{\;\;\nu}\wedge\ell^\nu$. 
The emergence of this torsion is 
spontaneous because it is a direct consequence and effective remnant 
of the original dynamics in 
${\cal B}^4_{\rm T}$
(the full calculation of  $\bar{T}^\mu$
will be given elsewhere). In any case, with or without effective torsion,  
the thus-emerging effective stress-energy tensor in 
$\bar{\cal B}^4_{\rm T}$ 
can be simply found via (\ref{avriemann}) from the Ricci tensor
$\bar R_{\mu\nu}=\kappa'^2\bar T_{\mu\nu}$, in terms of a
Planck-scale $\kappa'$ length parameter.

\section{Finite Taub strings  and the ${\cal B}^4_{\rm M}$, $\bar{\cal B}^4_{\rm M}$ cases}

\no
In the second half of this section we will have to lean on speculation,
with the mathematical rigor accordingly reduced. Finite  Taub strings,
open or closed, with  any number $n$  of ${\cal T}$ elements,
can also be realized as
stable non-singular and geodesically complete 4D vacua.
If they involve a relatively small 
number of ${\cal T}$ elements (like $n=1,2,3,\dots$),
they can be viewed as elemental 4D
bits taken off the ${\cal B}^4_{\rm T}$ string vacuum.
Their geodesic completeness can be established either by
joining the free $\Sigma$ boundaries of that bit
to create  a closed string, or extend its geodesics to infinity
exactly as with the Taub-NUT completion.
Thus, each one of the closed-string $n=1,2,3,\dots$ cases 
closes upon itself with no   links to infinity as
$/{\cal T}\backslash$, $/{\cal T}\vee {\cal T}\backslash$, 
$/{\cal T}\vee {\cal T}\vee {\cal T}\backslash,\dots$, 
where the initial and final slashes stand for the two free $\Sigma$ boundaries
(as the two halves of a $\vee$) to be identified.
The corresponding  open-string cases involve the
{\small NUT}$\vee{\cal T}\vee${\small NUT}, 
{\small NUT}$\vee{\cal T}\vee{\cal T}\vee${\small NUT},
{\small NUT}$\vee {\cal T}\vee{\cal T}\vee{\cal T}\vee${\small NUT}, $\dots$
configurations, along with all the features and charges inherited
after the paradigm of the original Taub-NUT  completion \cite{haw}.
For variant types of   ${\cal B}^4_{\rm T}$ vacua,  we observe
that  junctions  across the 
round-$S^3$  SL sections  $du=0$
at the $b^2=c^2$ positions of isotropy might also be
possible (subject to stability prerequisites), now involving
the mentioned $\Theta_0$ normal 
with ${\cal K}=-d\Theta_0$.

The ${\cal B}^4_{\rm M}$ of the Misner-case,
viewable as a  4D string 
in the present context (but cf. also \cite{haw}),
could conceivably replace the prototype ${\cal B}^4_{\rm T}$  
(accordingly reduced to a sub-case), 
but one might also conjecture on  a reverse possibility, as we briefly will.
In any case,  ${\cal B}^4_{\rm M}$
may well carry fundamental physical content in its Planck-scale mixmaster dynamics. 
Its principal $(a,b,c)$  radii   behave more like 
variables which can violate Bell's inequalities, just like quantum mechanical ones, 
with severe restrictions on their causal
transforms (like Fourier expansions) and  in contrast to
the fully  Fourier-expansible periodic $(b,b,c)$ radii of ${\cal B}^4_{\rm T}$. 
In likewise sharp contrast to the  ${\cal B}^4_{\rm T}$, $\bar{\cal B}^4_{\rm T}$
pair would be any attempt to reconstruct the
dynamics of  ${\cal B}^4_{\rm M}$ from excitations over  $\bar{\cal B}^4_{\rm M}$.
The latter has to be a  (3+3) type,
due to the overall equal participation of its $a,b,c$ radii, with
$d\bar{s}^2=\eta_{\mu\nu}\ell^\mu\ell^\nu$, easily calculable
$\bar{ R}^{\mu}_{\;\;\nu}$ curvature 2-form
and  effective stress-energy tensor  from $\bar R_{\mu\nu}=\kappa'^2\bar T_{\mu\nu}$.
However, the study of 
${\cal B}^4_{\rm M}$ itself is an entirely different enterprise, which might be 
facilitated  by means of
the probabilistic  involvement of local Taub breathing-mode dynamics\footnote{
As numerical  findings  indicate \cite{pier}, 
the $(a,b,c)$  principal radii 
in ${\cal B}^4_{\rm M}$ seem to form  patterns of
Taub  pairings, 
like a $(b,b,c)$ swaying after a while
to a  $(b,c,b)$, 
and so on. If so,  ${\cal B}^4_{\rm T}$ could 
help illuminate Planck-scale mixmaster 
dynamics in  the sense that the entire
time-evolution of a sufficiently small neighborhood in a homogeneous 
section of  ${\cal B}^4_{\rm M}$ could be statistically simulated
by $(b,b,c)$, $(b,c,b)$, or $(c,b,b)$ pairings
in an accordingly small neighborhood of the corresponding
${\cal B}^4_{\rm T}$ and with equally-shared overall participation.}. 
To further illustrate the point with a related aspect, we recall 
the mentioned  `loss of memory' peculiarity
with s-c geodesics, which allows the violation of
classical axioms with, e.g., bifurcations
in ${\cal B}^4_{\rm T}$  (to grow much worse  in 
${\cal B}^4_{\rm M}$).  These, combined with
the mentioned presence
of geodesics  trapped in the potential wells,
involve an  energy uncertainty 
$\delta E\sim|V_o|\geq \sqrt{2}/B$ from the s-c depth of those wells. This
uncertainty, multiplied by the conjugate one from the ${\delta u}=\sqrt{2}{\rm L_o}/B$
width of the wells (our `quantum of time'),  produces
a  phase-space surface element of  $({\delta E})({\delta u})\geq 2\rm L_o/B^2$.
Normalized as $({\delta E})({\delta u})\geq \frac{1}{2}$, this result fixes
the time scaling with $B=2$, hence with  ${\rm L'_o}={\rm L_o}$ as in Fig.\ref{fig}, in 
geometrized ${\rm L_o}=1$  units. 

\section{Discussion}

\no
The fundamental scale of any infinitely-long string can be identified  as 
larger or smaller, pending on physical interpretation,
as with the ${\rm L_o}$ scale of ${\cal B}^4_{\rm T}$ which is
equally-well identifiable with
Planck length or the Hubble radius ${\rm H_o}$, but {\em that} has
little or nothing to do with hierarchy. 
However, due to
the presence and physical content of a second scale
in ${\cal B}^4_{\rm T}$, the $\kappa_{\rm o}$,
we also have a natural hierarchy involved,
in the following sense. The scale of $\kappa_{\rm o}$
cannot significantly exceed  Planck length,
regardless of  the value of ${\rm L_o}$, because,
according to (\ref{stress}), the string tension would then weaken 
to the point of making ${\cal B}^4_{\rm T}$
unstable against large excitations.
Thus, $\kappa_{\rm o}$ can  
render ${\rm L_o}$  largest
as a Hubble radius ${\rm H_o}\gg\kappa_{\rm o}$,
which would place ${\cal B}^4_{\rm T}$ 
in the rank of a cosmological model,
or smallest 
as $\kappa_{\rm o}\sim{\rm L_o}$, which  
would render  ${\cal B}^4_{\rm T}$
pertinent to Planck-scale dynamics.
In both cases,  
stability and hierarchy  stem from the underlying
structure of  the quantized flow of time, established by the transition
of $N\in\IR\rightarrow n\in Z$. However,  
for Planck-length ${\rm L_ o}$, we also have the availability
of `de-quantization' or averaging by going to
the reverse limit  $n\in Z\rightarrow N\in\IR$, as seen.
By this we have not, of course, been taken  back to the  ${\cal T}$ we
started from, but to the static $\bar{\cal B}^4_{\rm T}$
and its effective enrichment. With $\bar{\cal B}^4_{\rm T}$ as 
ground state, excitations thereof must involve
two new  independent scales,
$\kappa$ and ${\rm L_1}\gg{\rm L_o}$,
in addition to $\kappa_{\rm o},\;{\rm L_o}$.
We will now briefly expand on some  of these aspects.

By the results in section 2, the stability  in ${\cal B}^4_{\rm T}$ 
relates to the potential-well  dynamics from $V(u)$
in-between the $\Sigma_n$. The  extension of this stability across the $\Sigma_n$
has emerged from aspects  which have sequentially  identified the $\Sigma_n$ 
as  boundaries of a (cured) physical singularity, as
junctions  and as
a gravitational shock-wave front, all relating to the fundamental
$\kappa_{\rm o}$-scale string tension on the (0,3) plane
along the null $\partial_u,L_3$  or
the  non-null $E_0$, $E_3$ vectors.  The `coincidence'  of 
identical components  (\ref{stress}) in so diverse
frames  is simply due to the  restrictive aspects of a  traceless
$S_{\mu\nu}$  layer on the (0,3) plane. The physical
content of this $S_{\mu\nu}$, essentially involved in qualifying 
${\cal B}^4_{\rm T}$ as a 4D string,
must be bestowed upon the string tension 
as a primitive geometric notion. 
It must be  accordingly 
transfered to the effective torsion $\bar T^\mu$, along with
the novel type of physical interpretation of the latter  in $\bar{\cal B}^4_{\rm T}$.

The propagating $(\partial_u,L_3)$  planes must all  have the 
correct orientation, because that relates to
the direction of propagation of the shock wave 
through the oriented $\Sigma_n$ surfaces. In turn, the sign of the normal
${\cal N}\!=\!L_3\!=\!c\Theta_3$ to the $\Sigma_n$ relates to the direction 
of propagation of time and it is carried over, via 
\mbox{$d{\cal N}\sim{\cal K}\sim (c^2)\dot{}$}, 
to the correct sign of the $S_{\mu\nu}$ stress-energy layers in (\ref{stress})
for stability. Reversal of that orientation, even at one junction,
would change that sign and apparently destroy the stability of the entire manifold,
so the underlying invariance is obviously a fundamental one.
The so resulting direction of time is in apparent accord with its other
fundamental aspects, namely its quantized flow at Planck scale and
its non-holonomic  involvement beyond that. The latter emerges  in relation to
the ($e^\mu$, $E_\nu$) non-singular orthonormal frames in (\ref{Frames}),
and could forbid time reversal for processes involving lengths and times 
of sufficiently larger-than-Planck scales.

Elements of the mentioned inter-dependences  survive in $\bar{\cal B}^4_{\rm T}$,
and they can be expected to also survive 
beyond the vacuum geometry.
By the holonomy theorems and the 
Cartan structure equations for any ${\cal R}^M_{\;\;N}$, $T^M$ set,
the  scale of torsion is completely independent from the scale of the 
Riemannian part $ R^M_{\;\;N}$ of the curvature \cite{traut}, 
so excitations off that geometry 
must be of the Palatini type (variations of metric
{\em plus} connection, independently). As a result, they must
necessarily involve,   beyond  $\kappa_{\rm o}$ and
${\rm L_o}$, the  gravitational coupling $\kappa$ 
from excitations of the metric and 
the independent  ${\rm L_1}$  from excitations of the connection. 
This can be applied  for the $SU(2)\times U(1)$ K-K manifold, 
calculated, not over  the conventional 
${\cal M}^4_{\rm o}\times S^3\times S^1$ ground state
(which is not stable and it is not even Ricci flat),  but over
$\bar{\cal B}^4_{\rm M}\times\IR^3\times S^1$.
Thus, the basic difference with the latter is stability
and the presence of the effective torsion $\bar T^\mu$, which 
makes it  compatible with the effective  vacuum equations.
The result uncovers the higgsless emergence of the correct
EW gauge-boson mass term from variation of the connection, whereby
${\rm L_1}$ can be  identified with the electroweak scale \cite{bat}.

Relating to issues on 
cosmic dynamics and  mixing
(such as isotropization, homogenization  and the horizon problem)
in relativistic astrophysics and cosmology, 
we may re-use (\ref{taub}), now
with ${\rm L_o}$ taken as a Hubble radius ${\rm H_o}$.  ${\cal B}^4_{\rm T}$ 
is then identified as an empty cosmological  model  in breathing mode
which  re-generates itself eternally. Likewise, the static
$\bar{\cal B}^4_{\rm T}$ and
$\bar{\cal B}^4_{\rm M}$  are non-singular 
empty universes with  SL sections, which are 
squashed or round $S^3$. The torsion therein  
should now be viewed as primordial rather than effective.
The effective stress-energy content will be again involved  as
$\bar R_{\mu\nu}=\kappa'^2\bar T_{\mu\nu}$, now
with $\kappa'=\kappa$. These models are stable against perturbation and
may thus relate to
galactic dynamics and to the  dark-matter/dark-energy  content of the universe.

\vskip .05in

I am grateful to A.~A.~Kehagias for illuminating discussions.

\end{document}